\documentclass[amssymb,aps,prd,nofootinbib]{revtex4-2}
\usepackage{graphicx}
\usepackage{subcaption} 
\usepackage{mathtools}
\usepackage{url}
\usepackage[normalem]{ulem}
\usepackage{caption}
\usepackage{enumitem}

\usepackage[colorlinks=true, linkcolor=blue, citecolor=red, urlcolor=red]{hyperref}
\newcommand{\be}{\begin{equation}}
	\newcommand{\ee}{\end{equation}}
\newcommand{\bea}{\begin{eqnarray}}
	\newcommand{\eea}{\end{eqnarray}}
\newcommand{\ba}[1]{\begin{array}{#1}}
	\newcommand{\ea}{\end{array}}
\newcommand{\nn}{\nonumber}

\newcommand{\ep}{\epsilon}

\newcommand{\del}{\partial}
\newcommand{\al}{\alpha}

\newcommand{\na}{\nabla}
\newcommand{\D}{\Delta}
\newcommand{\Ep}{\mathcal{E}}
\newcommand{\orcid}[1]{\href{https://orcid.org/#1}{\includegraphics[width=8pt]
		{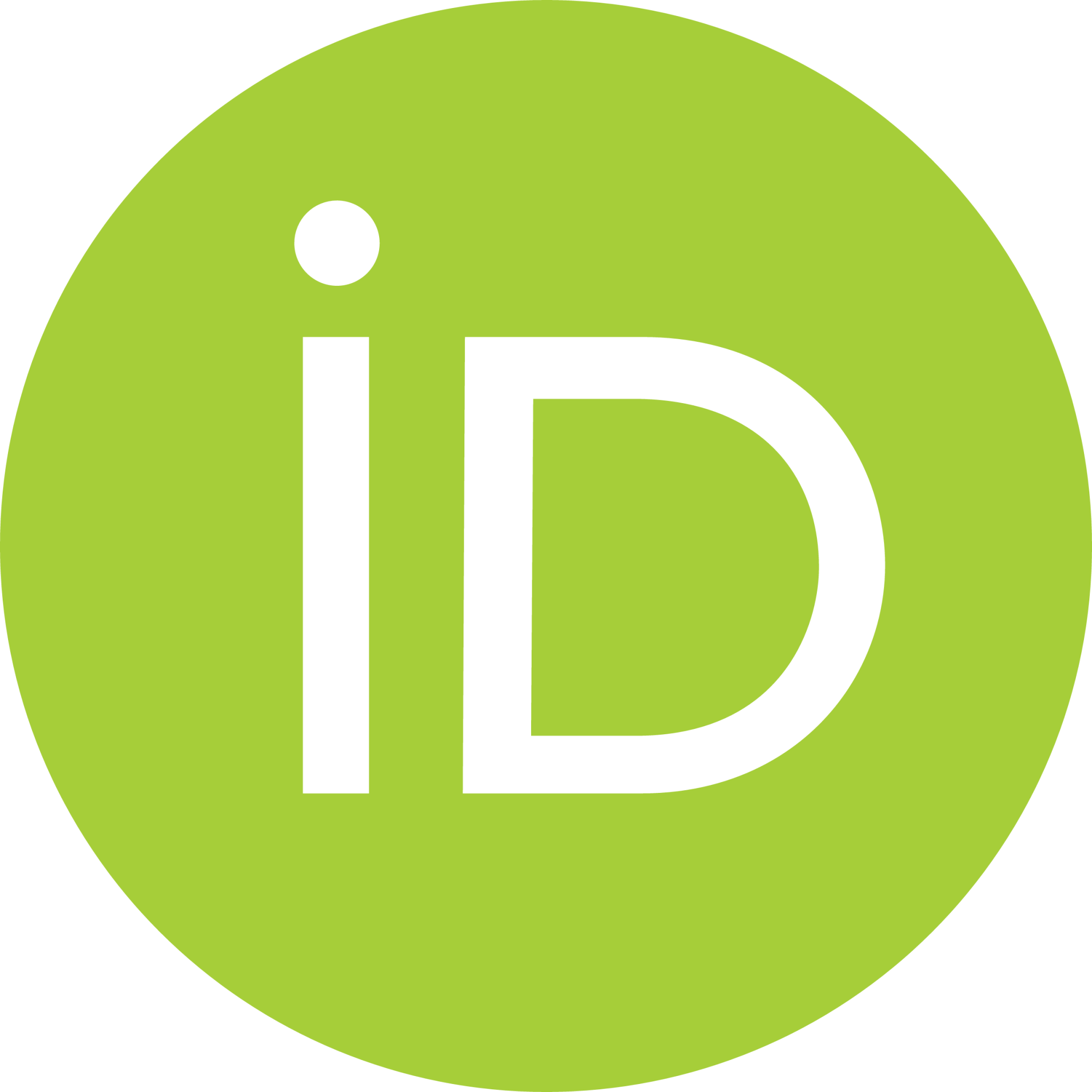}}}
\begin{document}
	\title{Electron Hydrodynamics in Graphene : Experimental and Theoretical Status}
	\author{Subhalaxmi Nayak\orcid{0009-0001-0145-6785}}
	\email{subhalaxmin@iitbhilai.ac.in}
	\author{Cho Win Aung\orcid{0000-0001-5684-2854}}
	\email{chowinaungnuclear@gmail.com}
	\author{Thandar Zaw Win\orcid{0009-0006-8033-3025}}
	\email{zawwin@iitbhilai.ac.in}
	\author{Ashutosh Dwibedi\orcid{0009-0004-1568-2806}}
	\email{ashutoshdwibedi92@gmail.com}
	\author{Sabyasachi Ghosh\orcid{0000-0003-1212-824X}}
	\email{sabyaphy@gmail.com}
	\author{Sesha Vempati\orcid{0000-0002-0536-7827}}
	\email{sesha@iitbhilai.ac.in}
	\affiliation{Department of Physics, Indian Institute of Technology Bhilai, Kutelabhata, Durg, 491002, Chhattisgarh, India}
	
	\begin{abstract}
	The present work comprehensively reviews electron hydrodynamics in graphene, highlighting both experimental observations and theoretical developments. Key experimental signatures such as negative vicinity resistance, Poiseuille flow, and significant violation of the Wiedemann-Franz (WF) law have been discussed, with special emphasis on Lorenz ratio measurements. In the theoretical direction, recent efforts have focused on developing hydrodynamic frameworks for calculating the thermodynamic and transport coefficients of electrons in graphene. The present work has briefly addressed the theoretical framework adopted by our group.
	\end{abstract}
	
	\maketitle
	
	\section{Introduction}
	 Carrier transport in solid state systems depends on various microscopic interactions such as scattering with electrons, phonons, impurities (charged or neutral), boundaries, disorder $etc.$~\cite{https://doi.org/10.1002/andp.201700043}. The scattering length or mean free path of charge carriers with electrons and phonons is inherently dependent on the carrier concentration and temperature. On the other hand, boundary scattering is associated with the size of the sample  $(W)$ and carrier confinement effects, if any. The net effect of transport is a result of the interplay of these processes \cite{https://doi.org/10.1002/andp.201700043,Lucasfong2018,Narozhny2019uib,NarozhnyGornyi2021},
	where the mean free path of the carrier is one of the key characteristics. Notably, the mean free path can be classified into two classes depending on whether the momentum is conserved or not. $i.e.$ momentum-conserving $(l_{MC})$ and relaxing $(l_{MR})$. The relative magnitude of the $l_{MC}$ and $l_{MR}$ with reference to $W$ would determine the type of carrier transport. This was initially predicted by Gurzhi  \cite{gurzhi1963minimum,gurzhi1965some,gurzhi_1968hydrodynamic}, where ballistic, diffusive, and hydrodynamic behaviors were noted. Specifically, hydrodynamics behavior is observed in different systems like phonons \cite{gurzhi_1968hydrodynamic,Ghosh_2022}, magnons \cite{PhysRev.188.898} at low temperature and by electrons in various materials such as $GaAs$ \cite{PhysRevB.51.13389,MOLENKAMP1994551}, $PdCoO_2$ \cite{Moll_2016}, $WP_2$ \cite{Gooth2018-am} high-mobility electrons in conductors \cite{doi:10.1126/science.aat8687}, cold atom \cite{PhysRevA.86.033614}, quark-gluon plasma (QGP) \cite{doi:10.1142/S0218301310014613,Palni:2024wdy}, and graphene \cite{	doi:10.1126/science.aad0201,crossno2016observation,Krishna_Kumar_2017} $etc$. The case with graphene is very special because of the linearly dependent energy dispersion, consequently yielding a “Dirac cone” in the band structure. Furthermore, graphene offers two-dimensional carrier confinement apart from tunable size and carrier densities \cite{neto2009electronic,RevModPhys.83.407}, making it an ideal platform for investigating the electron hydrodynamics and associated effects on the thermal as well as electrical properties. When the carrier density or chemical potential $(\mu)$ at finite temperature $T$ is altered $via$ doping, the carrier transport is very effectively controllable. Away from the Dirac point and when $\mu/k_B T >> 1$ ($k_B$ is Boltzmann’s constant), Fermi liquid behavior is observed as shown by conventional metals \cite{Landau1}. Near the Dirac point and when $\mu/k_B T << 1$, we observe the Dirac fluid-like behavior \cite{Lucasfong2018}. Also, within the hydrodynamic regime, electron and hole puddles are observed near the Dirac point \cite{PhysRevLett.116.126804}.
	\begin{figure*}  
		\centering 
		~ ~\includegraphics[scale=0.6]{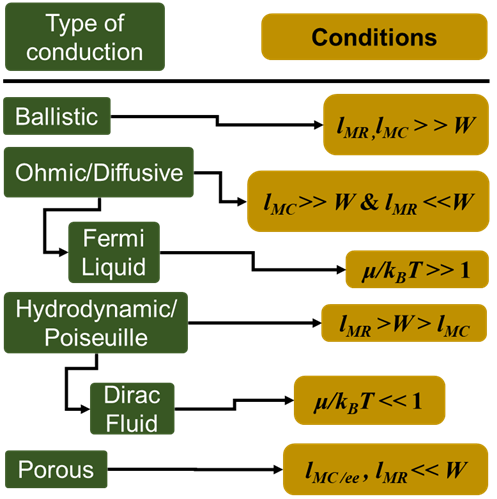} 
		\caption{ The type of charge conduction juxtaposed with that of momentum relaxing and conserving mean free paths. } 	\label{fig:1}
	\end{figure*} 
	Fig.(\ref{fig:1}) shows the detailed conditions in terms of $l_{MR},l_{MC},W,\mu$ and $T$ for various types of carrier transports, including the hydrodynamic flow of charge carriers. It can also be seen that the magnitude of carrier-carrier interaction is predominant in the Dirac fluid regime than that of the Fermi liquid \cite{Block2021-xx}.

	More interestingly, the relatively stronger interaction within the charge carriers in the hydrodynamic region impacts thermal and electrical transport properties. The electron hydrodynamics attracted both the experimental and theoretical aspects in a wide range of materials. In the experimental aspect, the violation of Wiedemann-Franz (WF) law \cite{crossno2016observation,PhysRevX.3.041008,Majumdar2025}, Mott law \cite{PhysRevLett.102.166808,PhysRevLett.102.096807,PhysRevLett.116.136802}, Poiseuille flow of electrons \cite{sulpizio2019visualizing}, holes \cite{crossno2016observation} were observed. In the theoretical aspect, various studies have appeared describing the hydrodynamic flow of electrons and its consequences on the thermodynamic and transport properties \cite{AnLucas2016,Dwibedi2025,win2025graphene,win2024wied,tu2023wiedemann,jaiswal2024spinhydrodynamics}. 
	Theoretical works \cite{Aung:2023vrr,PhysRevLett.103.025301} indicates shear viscosity to entropy density ratio, close to its  Kovtun-Son-Starinets (KSS) \cite{kovtun2005viscosity}  bound. Here, we present a review of the literature that describes the experimental observations and associated theories to explain the anomalies while confining ourselves to works related to graphene.
	
	\section{Experimental Signatures}

	In his pioneering work Gurzhi \cite{gurzhi1963minimum} considered, the electron flow through a thin wire with two typical scattering processes $viz$~ $(i)$ boundary (b) and $(ii)$ electron (e), impurity (i), or phonon (p) scattering.
	\begin{figure*}  
		\centering 
		~ ~\includegraphics[scale=0.7]{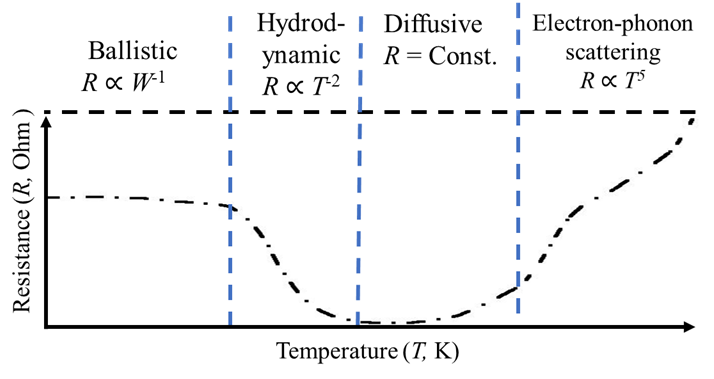} 
		\caption{ Variation of resistance against temperature shown schematically based on~\cite{gurzhi1963minimum,Narozhny_2022}. } 	\label{fig:2}
	\end{figure*} 
	In Fig.(\ref{fig:2}), we schematized the variation of resistance with respect to temperature as predicted in Ref.~\cite{gurzhi1963minimum}. Gurzhi observed that at low temperature $l_{MC}=l_{ee}>>W$, the boundary scattering is dominated, and the resistance is  $R\sim1/W$ which has no temperature dependence. This domain is basically the ballistic domain. As the temperature increases up to a certain threshold the momentum conserving electron-electron interaction takes over with $l_{MC}=l_{ee}< W$. This domain is governed by the hydrodynamic transport with the resistance of the sample falling as $T^{-2}$. A further increase in temperature drives the system into the diffusive regime where the momentum relaxing electron-impurity scattering predominates over electron-phonon scatterings. The resistance in this temperature range remains effectively constant until the electron-phonon interaction dominates and resistance increases as $T^{5}$. The hydrodynamics of electrons predicted by Gurzhi for a possible temperature range is seen in the experiments after a long time because of the challenges in making ultra clean samples. In this regard graphene turn out as a useful material of exploring the possibilities of electron hydrodynamics along with much more spectacular properties which we will explore in next paragraphs.
	
	Researchers are interested in graphene because of its potential to cause novel transport phenomena. As we know, graphene has high mobility and extraordinary electrical and thermal conductivities \cite{neto2009electronic}. Despite all the interesting properties, it also exhibits electronic hydrodynamic behavior \cite{https://doi.org/10.1002/andp.201700043}. This is observed in relatively pure samples
	at an intermediate temperature where the electron-phonon scattering is relatively weak.
	Later negative vicinity resistance \cite{doi:10.1126/science.aad0201}, WF law violation in graphene \cite{crossno2016observation}, super ballistic flow of the viscous fluid  through graphene constrictions \cite{Krishna_Kumar_2017} are observed. Negative vicinity or local resistance appears to be a crucial tool for observing vorticity in a confined system where viscous flow is shown. The existence of viscous flow is realized by vorticity, which may be observed by measuring the vicinity resistance  \cite{doi:10.1126/science.aad0201} by applying electrical leads close to each other in the vortex. Despite this, negative vicinity resistance may not be an appropriate tool to detect the hydrodynamic domain because ballistic systems also depict negative vicinity resistance both from theoretical and experimental perspectives \cite{Narozhny_2022}. The transformation from ballistic to hydrodynamic flow is observed by Sulpizio $et~ al.$ \cite{sulpizio2019visualizing}, in which the authors visualized Poiseuille's flow from the Hall field profile of the electron in the graphene sample. Using scanning carbon nanotube single-electron transistor, the spatial image of Poiseuille flow of electron moving through high-mobility graphene/hexagonal boron nitride (hBN) channels ( length $l=15~ \mu m$ and width $W=4.7~ \mu m$) were mapped. Hall field shows the transformation from the traditional ballistic flow shown by electrons in the normal metal to the parabolic or ideal Poiseuille's flow.
	\begin{figure*}  
		\centering 
		\includegraphics[scale=0.66]{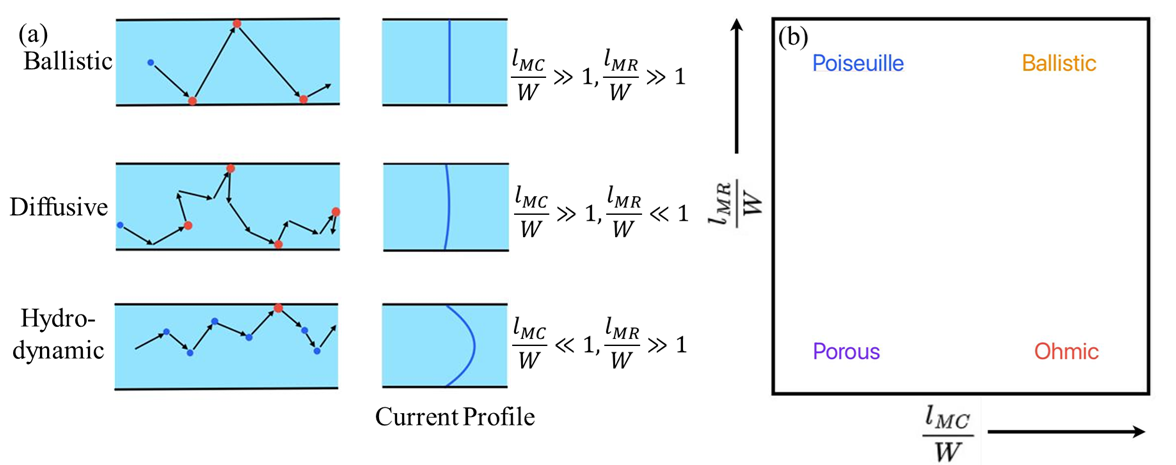}
		\caption{(a): Different scattering processes \cite{gurzhi1963minimum}  showing the current profile (b): Phase diagram showing different domains based on scattering process \cite{sulpizio2019visualizing}.} 	\label{fig:3}
	\end{figure*} 
	The different flow regimes--ballistic, Ohmic, porous, and Poiseuille, are governed by the mean free paths set by electron-electron interaction and electron-impurity, phonon, and disorder scatterings, which are illustrated in Fig.~(\ref{fig:3}).
	
	The signature of hydrodynamics can be understood from various experiments that suggest a breakdown of Fermi liquid paradigm. For instance, Lorenz ratio $L=\frac{\kappa}{\sigma T}$ becomes a constant $L_{0}=1.3 \times \frac{\pi^2}{3} (\frac{k_B}{e})^2$ in the usual Fermi liquid domain of metals, which is known as the WF law. Here, $\kappa$ represents the thermal conductivity, $\sigma$ denotes the electrical conductivity, and $T$ corresponds to the temperature of the system. The violation of the WF law ($L\neq L_{0}$) for a clean graphene could indicate a hydrodynamic flow of electrons and holes near the charge neutrality point (Dirac point). Experimentally the violation of WF law has been observed by measuring bipolar thermal conductivity with the consideration of diffusion of electrons and coupling between electrons and phonons \cite{PhysRevX.3.041008}. An increasing value of the Lorenz number is obtained in Ref.~\cite{PhysRevX.3.041008} where the system is predicted to be in a quantum critical regime near the Dirac point. Crossno $et~ al.$ \cite{crossno2016observation} worked on this using Johnson noise thermometry (JNT) with monolayer graphene samples encapsulated within hBN. They found that in higher chemical potential and intermediate temperature domain, the estimated Lorenz ratio satisfies the conventional value, but towards the low net carrier density domain or the charge neutrality point, a strong violation of the WF law is observed. The Lorenz ratio reached a higher value of around 22 at a temperature nearly equal to $60 ~K$. The huge violation of WF law near the Dirac point is attributed to the existence of the hydrodynamic behavior and revealing the presence of the Dirac fluid in graphene. Recently, the Ref.~\cite{Majumdar2025} has shown the WF law violation in ultra clean graphene samples with a very large value of the Lorenz ratio $L=400L_{0}$ near the Dirac point. The authors of Ref.~\cite{Majumdar2025} also observe a low shear viscosity ($\eta$) to entropy density ($s$) ratio whose value is close to the the holographic lower bound $i.e$., $\eta/ s\rightarrow \hbar/ 4 \pi k_B$. 
	
	\section{Theoretical Developments}\label{TD} 
	In this section, we will give a thorough overview of the thermodynamics and fluid dynamic transports of the charge carriers in graphene. The fluid dynamic regime and the transport coefficients like electrical conductivity, thermal conductivity, and shear viscosity in this regime are discussed in Sec.~\ref{Sec:FD}. The transport equations governing the dynamics of the charge carrier are also discussed in Sec.~\ref{Sec:FD}. Before that, we start the discussion with the non-fluid description usually followed in conventional metals and the possible validity of such descriptions in graphene in Sec.~\ref{Sec:NF}.
	
	\subsection{Non-fluid description}\label{Sec:NF}
	In the conventional metals where the  Fermi liquid theory is valid, the charge carriers are the weakly interacting electrons near the Fermi surface. The dispersion relation of the electrons is quadratic in momentum, and for an isotropic conduction band, one can write $E_{e}=\frac{p_{e}^{2}}{2m_{e}^{*}}$, where $m_{e}^{*}$ is the effective mass of the electrons in the bottom of the conduction band. The energy density of electrons for such a case can be written as,
	\begin{equation}
		\Ep_{e}=\frac{3 N_s }{2} \left(\frac{m_{e}^{*}}{2 \pi}\right)^{\frac{3}{2}} f_{\frac{5}{2}} (A)~ T^{\frac{5}{2}}  \label{NF1}
	\end{equation}
	where, $N_s$ is the spin degeneracy factor and  $f_{\frac{5}{2}} (A)$ is the Fermi integral whose general form is 
	\begin{equation}
		f_{\nu}(A) =\frac{1}{\Gamma(\nu)}\int\frac{x^{\nu-1}}{ A^{-1}e^x+1} dx. \label{E3}
	\end{equation} 
	$\nu$ takes all the integer and half-integer values and $A=e^\frac{\mu}{k_B T}$. The corresponding specific heat capacity can be calculated by using the definition $C_{e}=\frac{\del \Ep_{e}}{\del T}$. In metals, the limit $\mu=\ep_{F}\gg k_{B}T$ is well satisfied therefore one expresses $C_{e}$ using the Sommerfeld lemma result as, $C_{e}= \frac{\pi^2}{2} \frac{k_{B}^{2}T}{\mu}$. One then readily obtains the electrical and thermal conductivity as \cite{Ashcroft76},
	\bea
	&&\sigma_{e}=\frac{ne^2\tau_{c}}{m_{e}^{*}}=\frac{ne^2\lambda}{m_{e}^{*} v_F }~, \label{NF2}\\
	&&\kappa_{e}=\frac{1}{3} n v_F \lambda~ C_e~, \label{NF3}
	\eea 
	where $\lambda$, $v_{F}$ and $\tau_{c}=\lambda/v_{F}$ are the mean free path, Fermi velocity, and average collision time of the electrons respectively. Taking the ratio of $\kappa_{e}$ and $\sigma_{e}$ and using the relation $\mu=\ep_{F}=\frac{1}{2}m_{e}^{*}v_{F}^{2}$,  one obtains the usual Lorenz number,
	\begin{equation}
		\frac{\kappa_{e}}{\sigma_{e}T} =\frac{\pi^{2}}{3}\frac{k_B^{2}}{e^2}= L_0~.\label{NF4}
	\end{equation}
	In graphene, the conventional Fermi liquid theory is valid in the domain where $\mu \gg k_{B}T$. In this domain, the above theory may be expected to be valid with the following replacement 
	\begin{itemize}
		\item  the dispersion relation is changed to $E_{e}=p_{e}v_{F}$ with $v_{F}$ being the Fermi velocity of electrons in graphene
		\item the spatial dimension is reduced to two \cite{win2024wied,win2025graphene}. 
	\end{itemize}
	
	Most of the interesting features in graphene arise near the charge neutrality point $\mu\ll k_{B}T$, where a fluid dynamic description of the electrons is valid. Now, we move on to the next section to describe this interesting domain.
	\subsection{Fluid description}\label{Sec:FD}
	In graphene, by reducing the doping in cleaner samples, one reaches the Dirac fluid region  $\mu\ll k_{B}T$ where electron-electron momentum-conserving scattering dominates over the other momentum-relaxing scatterings. In this interaction-dominated regime, the hydrodynamics of the electrons are observed. In graphene, the relations: $\Vec{p}_{e,h}=\hbar\Vec{k}_{e,h}$, $E_{e,h}=(\hbar k_{e,h}) v_{F}$, and $\Vec{v}_{e,h}=v_{F}\frac{\Vec{k}_{e,h}}{k_{e,h}}$ are quite similar to the relations followed by a massless relativistic (ultra-relativistic) particle if we replace speed of light $c$ by electron Fermi velocity $v_{F}$. One may crudely assume the graphene $2$D system as an ultra-relativistic world with the highest speed $v_F$ (instead of $c$). Moreover, there exists numerous literature~\cite{RevModPhys.83.837,neto2009electronic} in the field of graphene where the relativistic behavior of electrons in graphene has been explored regarding $v_{F}$ as the limiting speed in the graphene world. So our analysis in this section has been borrowed from the principles of relativistic kinetic theory and relativistic fluid dynamics with a dimensional reduction $3 (\rm space)+1 (\rm time)$ to $2 (\rm space)+1 (\rm time)$ and the appropriate replacement of $c$ with $v_{F}$. Unless otherwise stated, any greek index ($\mu,\nu,\al,\beta$, $etc$.) runs from $0$ to $2$ whereas the latin indices ($i,j,k,l$, $etc$.) runs from $1$ to $2$. The Latin indices can be raised or lowered with the help of the Minkowski metric $\eta^{\mu\nu}=dia(1,-1,-1)$.

	In the Dirac fluid regime, the elementary degrees of freedom--electrons and holes, follow the microscopic energy-momentum conservation laws upon scattering, $i.e.$, $p^{\mu}_{1(e,h)}+p^{\mu}_{2(e,h)}=p^{\mu}_{3(e,h)}+p^{\mu}_{4(e,h)}$ for a collisional process: $1+2\xrightarrow{}3+4$. Here, we have combined the energy-momentum to form a lorentz-vector $p^{\mu}_{e,h}=(\frac{E_{e,h}}{v_{F}},p_{e,h}^{i})$. In the Dirac fluid regime, the microscopic conservation laws obeyed by the electrons and holes manifest themselves as conservation laws of an electron-hole fluid on a macroscopic level. The macroscopic conservation laws, which are direct consequences of microscopic conservation laws, can be written as follows,
	\begin{eqnarray}
		&&\text{energy-momentum conservation: }  p^{\mu}_{1(e,h)}+p^{\mu}_{2(e,h)}=p^{\mu}_{3(e,h)}+p^{\mu}_{4(e,h)}~, \implies \del_{\mu}T^{\mu\nu}=0~,\label{F1}\\
		&&\text{charge conservation: } N_{e}-N_{h}=\text{constant}~, \implies \del_{\mu}J^{\mu}=0~,\label{F2}
	\end{eqnarray}
	where $T^{\mu\nu}$ is the total stress-energy tensor of the electron-hole system, $N_{e}$ and $N_{h}$ are the total number of electrons and holes and $J^{\mu}$ is the current density. The space-time position $x^{\mu}$ and the space-time derivative $\del_{\mu}$ is defined as $x^{\mu}=(v_{F}t, x^{i})$  and $\del_{\mu}=\frac{\del}{\del x^{\mu}}=(v_{F}^{-1}\del_{t}, \del_{i})$.
	The meaning of the conservation equations becomes at once clear with the definitions of the different components of the stress-energy tensor and current density as,
	\begin{eqnarray}
		&&T^{00}=\Ep=\text{energy density of electron-hole fluid},~v_{F}T^{0i}=J_{E}^{i}=\text{energy flow in $i^{th}$ direction}~,\nn\\
		&& v_{F}^{-1} T^{i0}=\mathcal{P}^{i}=\text{momentum density},~ T^{ij}=\text{flow of $i^{th}$ component of momentum in $j^{th}$ direction}~,\nn\\
		&&v_{F}^{-1}J^{0}=\rho=\text{charge density},~ J^{i}=\text{charge current}~\label{F3}.
	\end{eqnarray}
	Using the definitions (\ref{F3}) in Eqs.~(\ref{F1}) and~(\ref{F2}), we can re-express the conservation laws in the familiar way as,
	\begin{eqnarray}
		&&\frac{\del \Ep}{\del t}=-\frac{\del J_{E}^{i}}{\del x^{i}}~,\label{F4}\\
		&& \frac{\del \mathcal{P}^{i}}{\del t}=-\frac{\del T^{ij}}{\del x^{j}}~,\label{F5}\\
		&& \frac{\del \rho}{\del t}=-\frac{\del J^{i}}{\del x^{i}}~,\label{F6}
	\end{eqnarray}
	where we can easily recognize Eq.~(\ref{F4}), Eq.~(\ref{F5}), and Eq.~(\ref{F6}) as energy conservation, momentum conservation, and charge conservation equations, respectively. Since all the conservation laws are contained in Eqs.~(\ref{F1}) and ~(\ref{F2}), everything boils down to the determination of the detailed structure of stress-energy tensor $T^{\mu\nu}$ and charge current $J^{\mu}$. The total stress-energy tensor and the net current density in a general out-of-equilibrium situation can be written as,
	\begin{eqnarray}
		&&T^{\mu\nu}=T_{0}^{\mu\nu} + T_{D}^{\mu\nu}~,\label{F7}\\
		&&J^{\mu}=J_{0}^{\mu}+J_{D}^{\mu}~,\label{F8}
	\end{eqnarray}
	where $T_{0}^{\mu\nu}$ and $J_{0}^{\mu}$ are the ideal part of the stress-energy tensor and current density with $T_{D}^{\mu\nu}$ and $J_{D}^{\mu}$ being the corresponding dissipative contributions. The ideal part of $T^{\mu\nu}$ and $J^{\mu}$ contain information about the thermodynamic variables of the electron-hole plasma and can be expressed as,
	\begin{eqnarray}
		&& T_{0}^{\mu\nu}= 4 v_{F}^{2}\bigg[\int \frac{d^{2}\Vec{p_e}}{h^{2}E_{e}}p^{\mu}_{e}p^{\nu}_{e}f^{0}_{e}+\int \frac{d^{2}\Vec{p}_{h}}{h^{2}E_{h}}p^{\mu}_{h}p^{\nu}_{h}f^{0}_{h} \bigg], \text{ or, }T_{0}^{\mu\nu}=v_{F}^{-2}~\Ep~ u^{\mu}u^{\nu}-P~\D^{\mu\nu}~,\label{F9}\\
		&& J_{0}^{\mu}=(-e) 4 v_{F}^{2}\bigg[\int \frac{d^{2}\Vec{p_e}}{h^{2}E_{e}}p^{\mu}_{e}f^{0}_{e}-\int \frac{d^{2}\Vec{p}_{h}}{h^{2}E_{h}}p^{\mu}_{h}f^{0}_{h} \bigg],\text{ or, }J_{0}^{\mu}=-en u^{\mu}\equiv \rho u^{\mu}~,\label{F10}
	\end{eqnarray}
	where $u^{\mu}=\gamma_{u}(v_{F},\vec{u})$ is the fluid velocity with $\gamma_{u}=1/\sqrt{1-u^{2}/v_{F}^{2}}$,  $\D^{\mu\nu}=\eta^{\mu\nu}-v_{F}^{-2}u^{\mu}u^{\nu}$. $\Ep$, $P$, and $n$ ($\rho$) are the total energy density, pressure, and net number density (charge density) of the system. The first microscopic set definitions of $T_{0}^{\mu\nu}$ and $J_{0}^{\mu}$ in terms of integrals of the distribution function of electrons $f_{e}=1/[e^{(p_{e}^{\mu}u_{\mu}-\mu)/k_{B}T}+1]$ and holes $f_{h}=1/[e^{(p_{h}^{\mu}u_{\mu}+\mu)/k_{B}T}+1]$ are the usual expressions of relativistic kinetic theory \cite{DeGroot:1980dk}. The degeneracy factor $4$ comes in the microscopic description because of the spin and valley degeneracy. The second macroscopic set of definitions of $T_{0}^{\mu\nu}$ and $J_{0}^{\mu}$ 
	have been arrived at by respecting the Lorentz invariance of the tensor structures \cite{jaiswal2016relativistic}. Comparing the microscopic and macroscopic definitions we have,
	\bea
	&& \rho= v_{F}^{-2}u_{\mu}J^{\mu}_{0}=-e(n_{e}-n_{h})=(-e)4\bigg[\int \frac{d^{2}\Vec{p}_{e}}{h^{2}} f^{0}_{e}-\int \frac{d^{2}\Vec{p}_{h}}{h^{2}} f^{0}_{h}\bigg]~,\label{F11}\\
	&& \Ep=v_{F}^{-2}u_{\mu}u_{\nu}T^{\mu\nu}_{0}=\Ep_{e}+\Ep_{h}= 4\bigg[\int \frac{d^{2}\Vec{p}_{e}}{h^{2}} E_{e} f^{0}_{e}+\int \frac{d^{2}\Vec{p}_{h}}{h^{2}} E_{h}f^{0}_{h}\bigg]~,\label{F12}\\
	&& P=-\frac{1}{2}\D_{\mu\nu}T_{0}^{\mu\nu}=P_{e}+ P_{h}=4v_{F}^{2}\bigg[\int \frac{d^{2}\Vec{p}_{e}}{h^{2}} \frac{p_{e}^{2}}{2E_{e}}  f^{0}_{e}+\int \frac{d^{2}\Vec{p}_{h}}{h^{2}} \frac{p_{h}^{2}}{2E_{h}}f^{0}_{h}\bigg]~,\label{F13}
	\eea 
	where $\Ep_{e(h)}$, $P_{e(h)}$ and $n_{e(h)}$ are the contribution of electrons (holes) to the total energy density, pressure, and net number density, respectively. Now, let us describe the dissipative parts of the stress-energy tensor and current density. In the absence of space-time gradients of fluid velocity $u^{\mu}$, chemical potential $\mu$, and temperature $T$, the system stays in equilibrium and the dissipative parts $J^{\mu}_{D}=T^{\mu\nu}_{D}=0$.  
	The system can be driven into out-of-equilibrium by creating the space-time gradients of either of the quantities--$u^{\mu}$, $\mu$, $T$, or by applying external electromagnetic fields. The macroscopic structure of the dissipation in a slight off-equilibrium scenario is expressed as--dissipative flows= transport coefficients $\times$ thermodynamic forces (or gradients). The transport coefficients can be evaluated by comparing the macroscopic relations provided in terms of thermodynamic forces with the microscopic kinetic theory descriptions. The microscopic and macroscopic expressions of the dissipative part of the stress-energy tensor and current density are given by,
	\begin{eqnarray}
		&& T_{D}^{\mu\nu}= 4 v_{F}^{2}\bigg[\int \frac{d^{2}\Vec{p_e}}{h^{2}E_{e}}p^{\mu}_{e}p^{\nu}_{e}\delta f_{e}+\int \frac{d^{2}\Vec{p}_{h}}{h^{2}E_{h}}p^{\mu}_{h}p^{\nu}_{h}\delta f_{h} \bigg], \text{ or, }T_{D}^{\mu\nu}=2 \eta~ \sigma^{\mu\nu}~,\label{F14}\\
		&& J_{D}^{\mu}=(-e) 4 v_{F}^{2}\bigg[\int \frac{d^{2}\Vec{p_e}}{h^{2}E_{e}}p^{\mu}_{e}\delta f_{e}-\int \frac{d^{2}\Vec{p}_{h}}{h^{2}E_{h}}p^{\mu}_{h}\delta f_{h} \bigg],\text{ or, }J_{D}^{\mu}=\sigma~\tilde{\Ep}^{\mu}+\sigma_{\kappa}\na^{\mu} T~,\label{F15}
	\end{eqnarray}
	where $\eta$, $\sigma$, and $\sigma_{\kappa}$ are, respectively, the shear viscosity, usual (diagonal ) conductivity, and off-diagonal conductivity of the system. The fluid velocity gradient $\sigma^{\mu\nu}$ and the thermodynamic gradient $\tilde{\mathcal{E}}^{\mu}$ containing the electric field $\tilde{E}^{\mu}$ are defined as:  $\sigma^{\mu\nu}\equiv\frac{1}{2}(\nabla^{\mu}u^{\nu}+\nabla^{\nu}u^{\mu})-\frac{1}{2}\D^{\mu\nu}\Theta$ and $\tilde{\Ep}^{\mu}\equiv \tilde{E}^{\mu}+\frac{1}{\rho}\na^{\mu}P$, where the spatial gradient $\nabla^{\mu}$ and expansion scalar $\Theta$ are defined as, $\nabla^{\mu}=\D^{\mu}_{\al}\del^{\al}$ and $\Theta=\nabla_{\al}u^{\al}$. $\tilde{E}^{\mu}$ is the covariant electric field defined through the help of the electromagnetic Faraday tensor $F^{\mu\nu}$ as $\tilde{E}^{\mu}=F^{\mu\nu}u_{\nu}$. The out-of-equilibrium part of the electron (hole) distribution function $\delta f_{e(h)}$ occurs in the microscopic description of the dissipative flows and can be found from the Boltzmann transport equation of electron (hole)~\cite{Dwibedi2025},
	\bea
	&& p^{\mu}_{e} \del_{\mu}f^{0}_{e} -\frac{e}{v_{F}^{2}}(\tilde{E}^{\mu}u^{\nu}-\tilde{E}^{\nu}u^{\mu})p_{e\nu}\frac{\del f^{0}_{e}}{\del p_{e}^{\mu}}=- \frac{u_{\mu}p^{\mu}_{e}}{v_{F}^{2}}\frac{\delta f_{e}}{\tau_{c}}~,\label{F16}\\
	&& p^{\mu}_{h} \del_{\mu}f^{0}_{h} +  \frac{e}{v_{F}^{2}}(\tilde{E}^{\mu}u^{\nu}-\tilde{E}^{\nu}u^{\mu})p_{h\nu}\frac{\del f^{0}_{h}}{\del p_{h}^{\mu}}=-\frac{u_{\mu}p^{\mu}_{h}}{v_{F}^{2}}\frac{\delta f_{h}}{\tau_{c}}~,\label{F17}
	\eea 
	where $\tau_{c}$ is the average collision time. Solving the above equations the $\delta f_{e,h}$ are obtained as,
	\begin{subequations} \label{F18}
		\begin{align}
			\delta f_{e} &= \frac{\tau_{c}v_{F}^{2}}{u_{\mu}p^{\mu}_{e}}\Bigg[\frac{p^{\al}_{e}p^{\beta}_{e}}{k_{B}T}\sigma_{\al\beta}-p^{\al}_{e}\bigg[\frac{n}{\Ep+P}u_{\beta}p^{\beta}_{e}-1\bigg]\left(-\na_{\al}\frac{\mu}{k_{B}T}+\frac{e\tilde{E}_{\al}}{k_{B}T} \right)\Bigg]f^{0}_{e}(1-f^{0}_{e})~,\\
			\delta f_{h} &= \frac{\tau_{c}v_{F}^{2}}{u_{\mu}p^{\mu}_{h}}\Bigg[\frac{p^{\al}_{h}p^{\beta}_{h}}{k_{B}T}\sigma_{\al\beta}-p^{\al}_{h}\bigg[\frac{n}{\Ep+P}u_{\beta}p^{\beta}_{h}+1\bigg]\left(-\na_{\al}\frac{\mu}{k_{B}T}+\frac{e\tilde{E}_{\al}}{k_{B}T} \right)\Bigg]f^{0}_{h}(1-f^{0}_{h})~.\
		\end{align}
	\end{subequations}
	The Eqs.~(\ref{F18}) give the microscopic way for the evaluation of $T^{\mu\nu}_{D}$ and $J^{\mu}_{D}$. We will first evaluate the dissipative part of the $J^{\mu}$ and then proceed to the evaluation of the dissipative part of $T^{\mu\nu}$. 
	Substituting Eqs.~(\ref{F18}) in the microscopic definition~(\ref{F15}) and using the Gibbs-Duhem relation: $n ~d\big(\frac{\mu}{k_{B}T}\big)=\frac{1}{k_{B}T}~dP+(\Ep+P)~d\big(\frac{1}{k_{B}T}\big)$ we have,
	\bea
	J^{\mu}&=&\frac{4e\pi\tau_{c}}{h^{2}}(k_{B}T)^{2}~\bigg[ 2(f_{2}(A^{-1})-f_{2}(A))+\frac{\Ep+P}{nk_{B}T} (f_{1}(A^{-1})+f_{1}(A))\bigg]\left[-\frac{\rho}{\Ep+P}\tilde{\Ep}^{\mu}+\frac{1}{T}\na^{\mu}T\right]~.\nn\\
	\label{F19-2}
	\eea 
	Comparing the macroscopic definition of $J^{\mu}_{D}$ given in Eq.~(\ref{F15}) with Eq.~(\ref{F19-2}) we have,
	\bea
	\sigma= 4\pi\tau_{c}e^{2}\left(\frac{k_{B}T}{h}\right)^{2}~\frac{n}{\Ep+P}\bigg[ 2(f_{2}(A^{-1})-f_{2}(A))+\frac{\Ep+P}{nk_{B}T} (f_{1}(A^{-1})+f_{1}(A))\bigg]~,\label{F21}\\
	\sigma_{\kappa}= \frac{4\pi\tau_{c}e}{T}\left(\frac{k_{B}T}{h}\right)^{2}~\bigg[ 2(f_{2}(A^{-1})-f_{2}(A))+\frac{\Ep+P}{nk_{B}T} (f_{1}(A^{-1})+f_{1}(A))\bigg]\label{F22}~.
	\eea  
	Another important dissipative flow related to charge flow is the heat flow. The heat flow $q^{\mu}$ for relativistic fluid is defined as the difference between the dissipative part of energy flow $W^{\mu}\equiv \Delta_{\al}^{\mu}T^{\al\beta}u_{\beta}$ and enthalpy flow $h^{\mu}\equiv -\frac{1}{e} \mathfrak{h} \Delta^{\mu}_{\al}J^{\al}_{D}$ \cite{DeGroot:1980dk} $i.e.$, $q^{\mu}=W^{\mu}-h^{\mu}=\Delta_{\al}^{\mu}(T^{\al\beta}u_{\beta}+\frac{\mathfrak{h}}{e}J^{\al}_{D})$, where $\mathfrak{h}\equiv \frac{\Ep+P}{n}$ is the enthalpy per particle. In the Landau-Lifshitz hydrodynamic frame where the dissipative part of energy flow vanishes \cite{DeGroot:1980dk} and the expression of heat flow becomes: $q^{\mu}=-h^{\mu}=-\frac{\Ep+P}{n}\Delta_{\al}^{\mu}N^{\al}_{D}=-(\Ep+P)\frac{J^{\mu}}{\rho}$. The preceding definition of heat flow with Eq.~(\ref{F19-2}) gives the following expression for $q^{\mu}$:
	\bea
	q^{\mu}&=&\frac{4\pi\tau_{c}}{h^{2}}(k_{B}T)^{2}\frac{\Ep+P}{n}~\bigg[ 2(f_{2}(A^{-1})-f_{2}(A))+\frac{\Ep+P}{nk_{B}T} (f_{1}(A^{-1})+f_{1}(A))\bigg]\left[-\frac{\rho}{\Ep+P}\tilde{\Ep}^{\mu}+\frac{1}{T}\na^{\mu}T\right]~.\nn\\
	\label{F23}
	\eea
	Identifying Eq.~(\ref{F23}) with the expression $q^{\mu}=\kappa_{\sigma}~\tilde{\Ep}^{\mu}+\kappa\na^{\mu} T$ we have,
	\bea
	&& \kappa_{\sigma}=4\pi \tau_{c} e \left(\frac{k_{B}T}{h}\right)^{2}\bigg[ 2(f_{2}(A^{-1})-f_{2}(A))+\frac{\Ep+P}{nk_{B}T} (f_{1}(A^{-1})+f_{1}(A))\bigg]~,\label{F24}\\
	&&\kappa=4\pi \tau_{c} k_{B}\left(\frac{k_{B}T}{h}\right)^{2}\frac{\Ep+P}{nk_{B}T}\bigg[ 2(f_{2}(A^{-1})-f_{2}(A))+\frac{\Ep+P}{nk_{B}T} (f_{1}(A^{-1})+f_{1}(A))\bigg]~\label{F25}.
	\eea
	The Lorenz ratio is calculated as follows,
	\bea
	&&L=\frac{\kappa}{\sigma T}=\left(\frac{\ep+P}{nk_{B}T}\right)^{2}\frac{k_{B}^{2}}{e^{2}}=\left(\frac{\mathfrak{h}}{k_{B}T}\right)^{2}\frac{k_{B}^{2}}{e^{2}}~.\label{F26}
	\eea
	At the end, let us tabulate the non-fluid and fluid-based expression of $\sigma$, $\kappa$, and $L$ in  Table- (\ref{TableTransport}). Readers can get a clear difference in the mathematical structures of those quantities for the non-fluid and fluid based frameworks of graphene.
	
	\begin{table}[h!]
		\centering
		\begin{tabular}{ |c|c|p{6.5cm}| } 
			\hline
			Transport Coefficients & Non-fluid & ~~~~~~~~~~~~~~~~~~~~~~~~Fluid \\ 		 
			\hline
			Electrical Conductivity $\sigma$ & $\frac{n e^2 \tau_c}{m_{e}^{*}}$ & $4\pi\tau_{c}e^{2}\left(\frac{k_{B}T}{h}\right)^{2} \frac{n}{\epsilon+P} \bigg[ 
			2(f_{2}(A^{-1}) - f_{2}(A)) 
			\newline ~~~~~~~~~~~~~~	+ \frac{\epsilon + P}{n k_{B} T} \big(f_{1}(A^{-1}) + f_{1}(A)\big)
			\bigg]$ \cite{Dwibedi2025} \\ 
			Thermal conductivity $\kappa$ & $\frac{1}{3} n v_F \lambda~ C_e$ & $4\pi \tau_{c} k_{B}\left(\frac{k_{B}T}{h}\right)^{2}\frac{\epsilon+P}{n k_{B} T} \bigg[ 
			2(f_{2}(A^{-1}) - f_{2}(A)) 
			\newline ~~~~~~~~~~~~~~	+ \frac{\epsilon + P}{n k_{B} T} \big(f_{1}(A^{-1}) + f_{1}(A)\big)
			\bigg]$ \cite{Dwibedi2025} \\ 
			Wiedemann-Franz law $\frac{L}{L_0}$ & $\frac{\pi^2}{3} \left(\frac{k_B}{e}\right)^2$ \cite{Ashcroft76} & $\left(\frac{\mathfrak{h}}{k_{B}T}\right)^{2} \frac{k_{B}^{2}}{e^{2}}$ \cite{Dwibedi2025} \\ 
			\hline
		\end{tabular} 
		\caption{Summary of Transport Coefficients in Non-fluid and fluid domains}
		\label{TableTransport}
	\end{table}
	
	Next, to calculate the shear viscosity coefficient, the dissipative part of $T^{\mu\nu}$ can be evaluated with the help of Eq.~(\ref{F14}) and Eqs.~(\ref{F18}) as,
	\bea
	&&T_{D}^{\mu\nu}=4 v_{F}^{2}\frac{\tau_{c}v_{F}^{2}}{h^{2}k_{B}T}\Bigg[\int \frac{p_{e}^{4}}{8E_{e}^{2}}f^{0}_{e}(1-f^{0}_{e})~d^{2}p_{e}+ \int \frac{p_{h}^{4}}{8E_{h}^{2}}f^{0}_{h}(1-f^{0}_{h})~d^{2}p_{h}\Bigg] ~2\sigma^{\mu\nu}\nn\\
	\implies && T_{D}^{\mu\nu}= 6 \frac{\pi \tau_{c}(k_{B}T)^{3}}{(hv_{F})^{2}} [f_{3}(A)+f_{3}(A^{-1})]  ~2\sigma^{\mu\nu}~.\label{F27}
	\eea
	Comparing Eq.~(\ref{F27}) with the macroscopic definitions of Eq.~(\ref{F14}) the shear viscosity $\eta$ is given by,
	\bea
	\eta= 6 \frac{\pi \tau_{c}(k_{B}T)^{3}}{(hv_{F})^{2}} \Bigg[f_{3}(A)+f_{3}(A^{-1})\Bigg]~. \label{F28}
	\eea 
	One interesting quantity that measures the fluidity of a system is the ratio between shear viscosity and entropy density. The entropy density in graphene can be obtained from Euler's thermodynamic relation $Ts=\Ep+P-\mu~n$. One can express $\eta/s$ for the electron-hole plasma in graphene in terms of Fermi integral functions as, 
	\bea
	&& \frac{\eta}{s}=\frac{3}{4}\tau_{c}T\frac{f_{3}(A)+f_{3}(A^{-1})}{3[f_{3}(A)+f_{3}(A^{-1})]-\frac{\mu}{k_{B}T}[f_{2}(A)-f_{2}(A^{-1})]}~.\label{F29}
	\eea  
	
	Readers can refer to Ref.~\cite{Aung:2023vrr} for the RTA-based formulation of the electronic contribution only to both $\eta$ and the ratio $\frac{\eta}{s}$. The present manuscript for the first time, reveals the detailed RTA-based mathematical structures of $\eta$  and $\frac{\eta}{s}$ for electron-hole plasma in graphene system in Eqs.~(\ref{F28}) and (\ref{F29}), respectively. 
	\section{Results and discussion}\label{Res}
	\begin{figure*}  
		\centering 
		\includegraphics[scale=0.48]{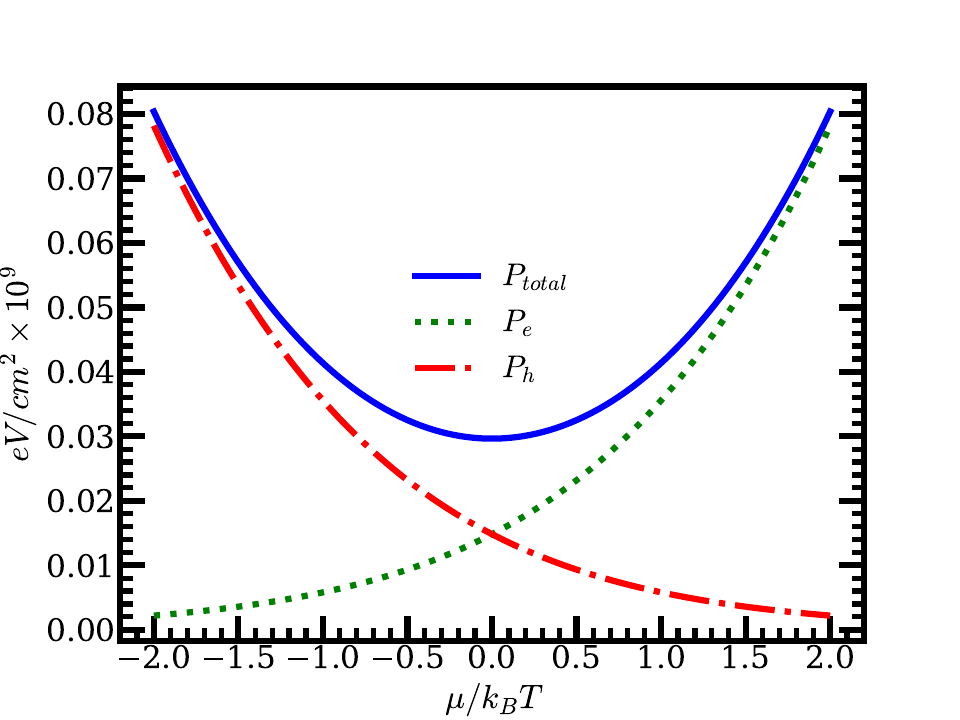}
		\hfill
		\includegraphics[scale=0.48]{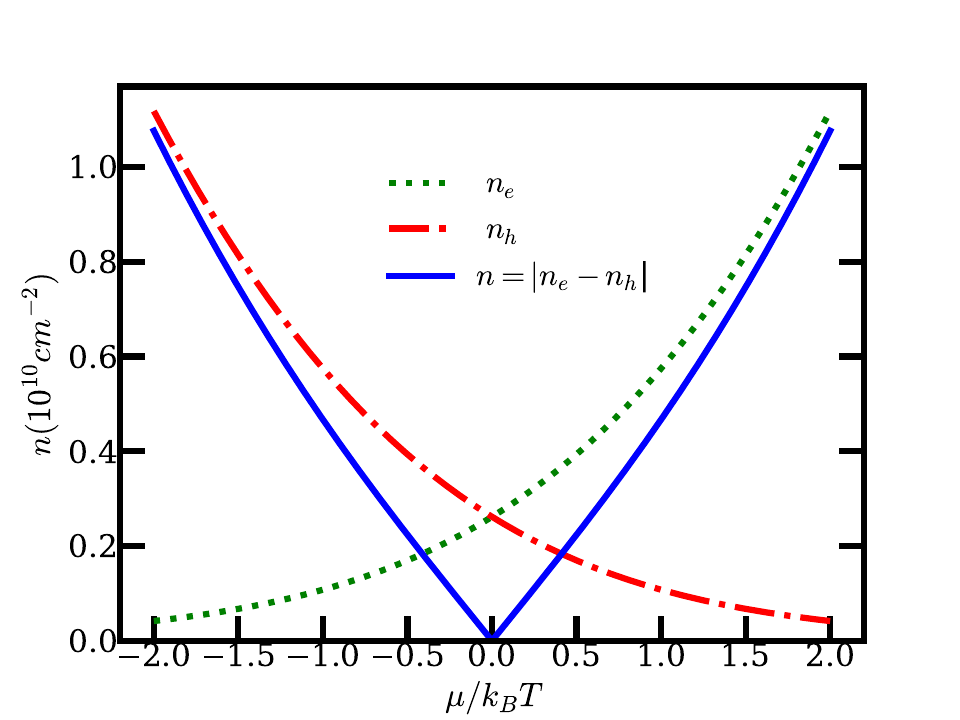}
		\caption{Pressure (Left panel) and number density  (Right panel) against  ($\frac{\mu }{k_B T}$) \cite{Dwibedi2025}.} 	\label{fig:thermo}
	\end{figure*}
	
	In this section, we display the variation of thermodynamic and transport variables obtained in Sec.~\ref{Sec:FD}. In the thermodynamic sector the Eqs.~(\ref{F11})--(\ref{F13}) are used to generate the results of net number density $n$, energy density $\Ep$ and pressure $P$. In the transport sector, Eqs.~(\ref{F26}) and~(\ref{F29}) are used to depict the result of Lorenz ratio normalized by $L_{0}$ and shear viscosity to entropy density ratio $\eta/s$. They are discussed one by one in the next paragraphs.

	In the left panel of Fig.~(\ref{fig:thermo}), we have shown the contribution of electrons and holes to the pressure as a function of $\frac{\mu}{k_B T}$ at a fixed $T=60$ K ($k_{B}T=5.14$ meV). Around the charge neutrality point and for $\mu$ value $-2.6$ to $2.6$ meV, the pressure for electron and hole shows an increasing value (blue line) while the individual contribution of electron (green dotted line) (hole (red dot-dashed line)) to the pressure increases for the increasing positive (decreasing negative) value of $\frac{\mu}{k_BT}$ respectively. A similar trend can be observed for energy density by using $\Ep = 2P$. We display the contribution of electron density (green dotted line) and hole density (red dot-dashed line) to the absolute value of the net number density (blue line) in the right panel of Fig.~(\ref{fig:thermo}). It is observed that at $T=60$ K, the carrier transport is dominated by electrons in the domain $\mu>10$ meV, whereas in the region $\mu<-10$ meV, the carrier transport is dominated by holes. For $-10<\mu<10$ meV, both the electrons and holes contribute significantly to the carrier transport. 
	\begin{figure*}  
		\centering 
		\includegraphics[scale=0.48]{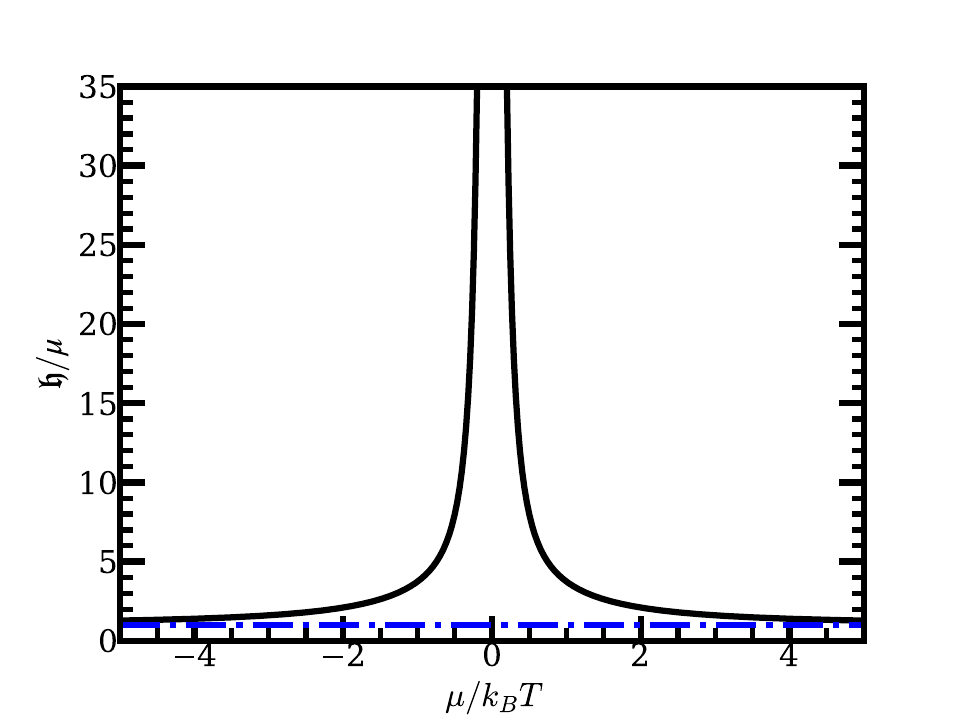}
		\hfill
		\includegraphics[scale=0.48]{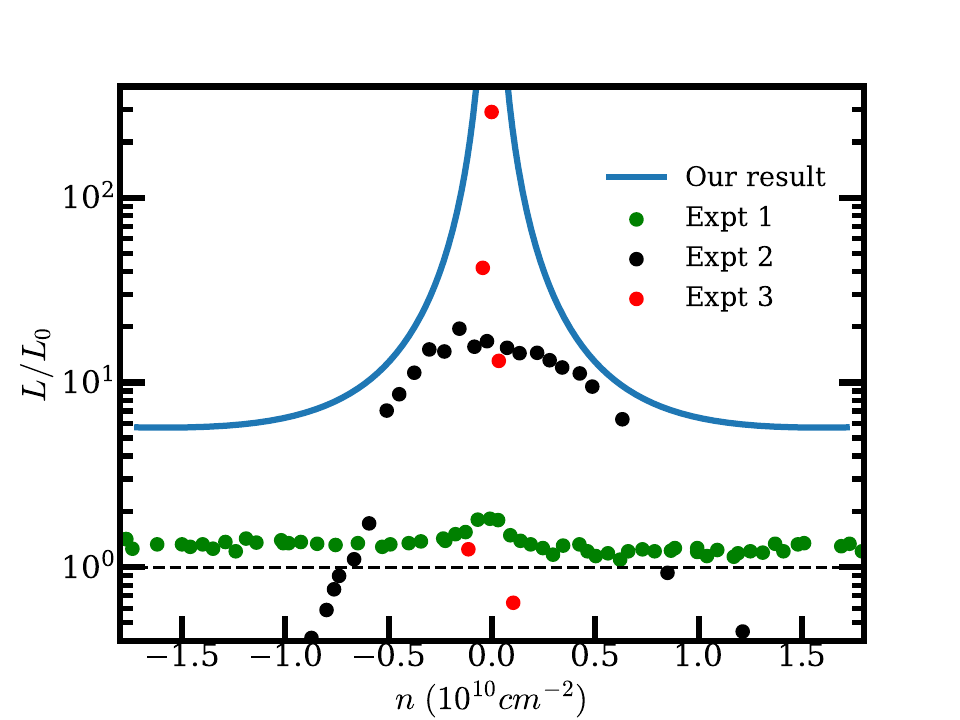}
		\caption{Enthalpy density per particle against ($\frac{\mu }{k_B T}$) \cite{Dwibedi2025} (Left panel) and Lorenz ratio vs number density $(n)$ with Experiment-1~\cite{PhysRevX.3.041008}, Experiment-2~\cite{crossno2016observation}, Experiment-3~\cite{Majumdar2025} and our result~\cite{Dwibedi2025} (Right panel).} \label{fig:lor}
	\end{figure*}
	
	In the left panel of Fig.~(\ref{fig:lor}), we have presented the variation of enthalpy density per particle scaled by $\mu$ against $\frac{\mu}{k_B T}$. It is worthy to point out that $\mathfrak{h}$ emerge as a crucial quantity in the hydrodynamic regime of transport as is evident from the expressions~(\ref{F21}), (\ref{F22}), and (\ref{F24})--(\ref{F26}). Here we can notice that in the Fermi liquid domain, $\frac{\mathfrak{h}}{\mu}$ tends to $1$ for $\frac{\mu}{k_B T}>3.5$ at a fixed temperature $T=60$ K and shows a deviation from one for $\frac{\mu}{k_B T}< 3.5$~. Near the Dirac or charge neutrality point for $\frac{\mu}{k_B T}<< 1$, it diverges. In the right panel of Fig.~(\ref{fig:lor}), we have performed a comparative study for the different results of the Lorenz ratio measured in different experimental works. Experiment-1 (green dots)~\cite{PhysRevX.3.041008} denotes the measured Lorenz ratio by observing the bipolar thermal conductivity and found the violation of WF law up to $1.3~ L_0$ whereas in Experiment-2 (black dots)~\cite{crossno2016observation} the Lorenz ratio is observed by using Johnson noise thermometry at $T=60$ K for ultra-pure graphene sample with an enhancement of Lorenz ratio up to $22~ L_0$. Experiment-3 (red dots)~\cite{Majumdar2025} depicts the violation of Wiedemann-Franz law at very low temperature $T= 19$ K, where the deviation of the Lorenz ratio from Fermi liquid value is found up to $400~ L_0$. The blue curve shows the value of Lorenz ratio~\cite{Dwibedi2025} according to Eq.~(\ref{F26}) where the enthalpy per particle is responsible for a huge violation of WF law near the Dirac point in the graphene system at a fixed temperature $T=60$ K.
	
	\begin{figure*}  
		\centering 
		\includegraphics[scale=0.5]{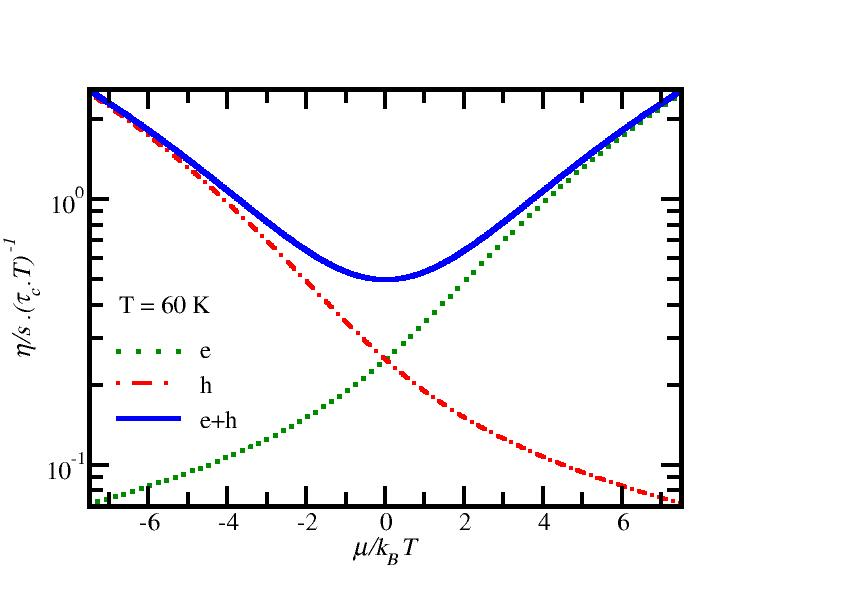}
		\caption{Shear viscosity to entropy density in terms of $\mu/k_B T$ .} 	\label{fig:eta_s}
	\end{figure*}
	
	In Fig.~(\ref{fig:eta_s}), we have shown the variation of normalized $\frac{\eta}{s}$ (blue solid line) with $\mu/k_BT $ as per Eq.~(\ref{F29}) where the contribution of electron (green dotted line), as well as holes (red dot-dashed line), are depicted at a fixed temperature $T=60$ K. As we have observed for the case of thermodynamic variables the electronic contribution to the $\eta/s$ becomes negligible for $\mu/k_{B}T<-2$ whereas the hole contribution becomes very small for $\mu/k_{B}T>2$. In the domain $-2<\mu/k_{B}T<2$, both electron and hole contribute significantly to the viscosity of the electron fluid in graphene. Interestingly, the $\eta/(s~\tau_{c}T)$ has a valley-shaped pattern with the minimum at $\mu=0$ corresponding to the charge neutrality point. Here, we have shown only the thermodynamic phase space part of  $\eta/s$ by normalizing $\eta/s$ by $\tau_{c}$. Near the charge neutrality, the phase space part shows a dip, but with the increasing value of $\frac{\mu}{k_B T}$, it increases and merges with the electronic contribution to the $\eta/s$. As we know, $\tau_{c}(T)$ carries all the information regarding the interactions and to determine the actual variation of $\eta/s$ one needs to calculate $\tau_{c}(T)$ incorporating all the scattering mechanisms. The numerical value $\eta/s$ serves as a fluidity measure and gives information about the interaction strength between the medium constituents. It is found to be very small for most of the quantum critical fluids and close to the lower bound limit in quark-gluon plasma domain~\cite{PhysRevLett.103.025301}. At room temperature, $\eta/s$ for graphene is observed to be four times larger than the KSS or holographic lower bound in the Ref.~\cite{Majumdar2025}.

	Near the charge neutrality point, the graphene is expected to behave as a quantum critical fluid and a quasi-relativistic plasma, which is better known as Dirac fluid~\cite{doi:10.1126/science.aat8687}. In the Dirac fluid domain, the electrons and holes strongly interact with each other. This fact initiates a new theoretical framework- electron hydrodynamics in graphene \cite{Narozhny_2022,Lucasfong2018,Narozhny2019uib,win2025graphene,win2024wied}. Instead of diffusion or Ohmic motion of electrons, hydrodynamic motion of electrons is observed in graphene for a particular low temperature and carrier density. Some interesting phenomena like Poiseuille's flow \cite{sulpizio2019visualizing}, negative vicinity resistance \cite{	doi:10.1126/science.aad0201}, Wiedemann-Franz law violation  \cite{crossno2016observation} $etc.$ are experimentally observed \cite{sulpizio2019visualizing,doi:10.1126/science.aad0201,crossno2016observation}, which can not be possible in Ohmic/diffusion motion of electrons. So, they can be considered as the signature of electron hydrodynamics. Among these experimentally observed signatures, theoretical groups have attempted to explain WF law violations. In Ref.~\cite{AnLucas2016}, the development of a non-perturbative relativistic hydrodynamic theory of electron transport in graphene fluid near a quantum critical point is addressed. This results in the violation of WF law and the theoretical derivations are found to be in agreement with the experimental data. Ref.~\cite{AnLucas2016} showed that quantum critical conductivity $\sigma_Q\approx \frac{e^2}{h}$ \cite{PhysRevB.76.144502} plays an important role in  WF law violation. Another work \cite{tu2023wiedemann} also shows the violation of WF law by applying Fermi liquid theory to the extrinsic doped graphene. They have established a theory to describe the work of \cite{crossno2016observation} by producing a gap near the Dirac point by the h-BN substrate using bipolar diffusion Boltzmann transport model involving disorder and phonon scattering. Ref.~\cite{rycerz2021wiedemann} has shown the violation of WF law by using Landauer–Büttiker formalism. In this context, our IIT Bhilai eHD group has presented a systematic formalism of electron hydrodynamic based electrical and thermal conductivity expressions and Lorenz ratio; we have also identified grossly the fluid and non-fluid domains along the carrier density axis at a fixed temperature and guessed the transition region from one domain to the other.
	\section{Summary}
	This paper presents a comprehensive study of electron hydrodynamics in graphene, where under specific conditions—particularly near the charge neutrality point—electrons behave collectively like a viscous fluid due to strong electron-electron interactions. This is evidenced by Poiseuille flow patterns and significant violations of the Wiedemann-Franz (WF) law, as seen in ultra-clean graphene samples. Theoretically, the paper contrasts conventional Fermi liquid behavior with the Dirac fluid regime using kinetic theory and quasi-relativistic hydrodynamic models, calculating thermodynamic quantities and transport coefficients. The Lorenz ratio is shown to deviate strongly from its classical value in the Dirac fluid regime, aligning with experimental findings. Additionally, graphene's shear viscosity to entropy density ratio may approach the universal lower bound known from holographic models, suggesting that graphene behaves as a nearly perfect fluid, similar to quark-gluon plasma in high-energy physics.
	
	\section{Acknowledgment}
	This work was supported in
	part by the Ministry of Education, Government of India  (S.N., A.D.), Doctoral fellowship in India, Ministry of education, Government of India (C.W.A., T.Z.W.), and Board of Research in Nuclear Sciences (BRNS) and Department of Atomic Energy (DAE), Govt. Of India with Grant Nos. 57/14/01/2024-BRNS/313 (S.G.).\\
	{\bf Funding: } This research received no external funding.\\
	{\bf Data Availability Statement:} 
	This study is purely theoretical and does not generate new datasets. The experimental data referenced for comparison are available in their respective original publications, as cited in the manuscript.\\
	{\bf Code Availability Statement:} This study does not involve custom computational code.\\
	\\
	{\bf Declarations}\\
	{\bf Conflict of interest:} The authors declare no conflicts of interest relevant to this study.

	\bibliographystyle{unsrturl}
	\bibliography{reference}

\end{document}